\documentclass[a4paper,oneside]{article}

\usepackage{IEEEtrantools}
\usepackage{amsfonts}
\usepackage{amsthm}
\usepackage{dansimath}
\usepackage[square,authoryear,comma]{natbib} 
\usepackage{graphicx}
\usepackage[english]{babel}
\usepackage{amsmath}
\usepackage{stfloats}
\usepackage{enumerate}
\usepackage{fullpage}
\usepackage{subcaption}

\begin{document}

\title{Robust MRAC augmentation of flight control laws\\ for center of gravity adaptation}
\author{Daniel Simon}
\date{\today}
\maketitle
\abstract{When an aircraft is flying and burning fuel the center of gravity (c.g.) of the aircraft shifts slowly. The c.g. can also be shifted abruptly when e.g. a fighter aircraft releases a weapon. The shift in c.g. is difficult to measure or estimate so the flight control systems need to be robustly designed to cope with this variation. However for fighter aircrafts with high manoeuvrability there is room for improvements. In this project we investigate if the use of adaptive control law augmentation can be used to better cope with the change in c.g. We augment a baseline controller with a robust Model Reference Adaptive Control (MRAC) design and analyse its benefits and possible issues.}

\section{Aircraft model and baseline controller}
\begin{figure}[hbt]
  \centering
  \includegraphics[width=0.3\textwidth]{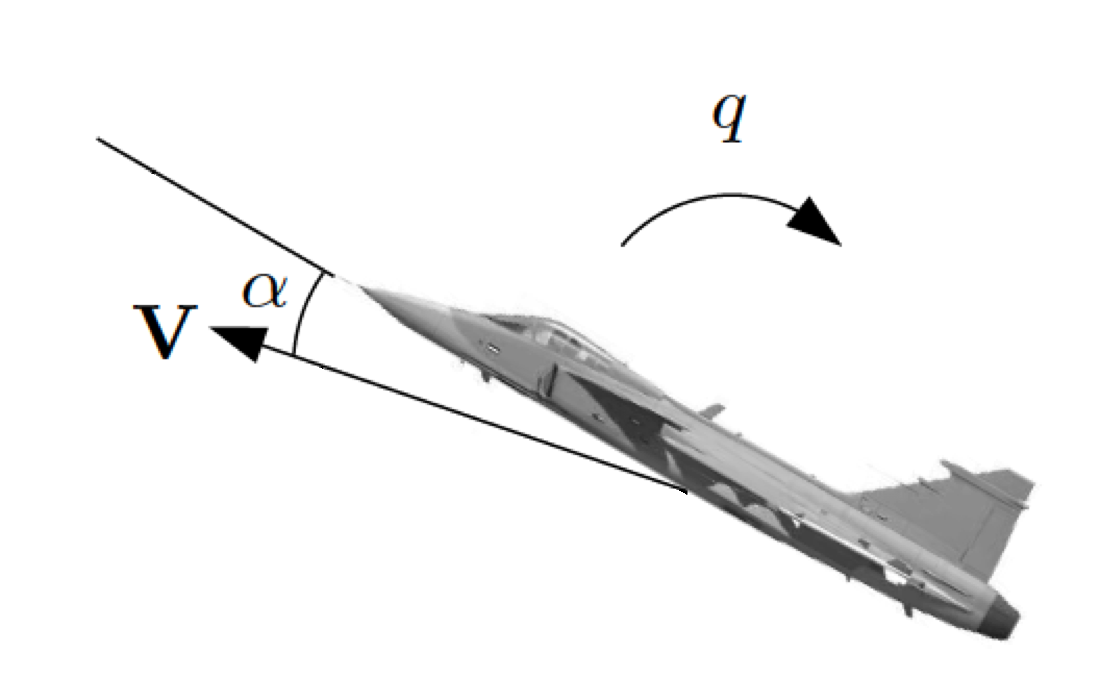}
  \caption{Definition of angles for aircraft control}
  \label{fig:aircraft}
\end{figure}

The dynamics that we will consider in this report is a linearized version of the pitch dynamics of the ADMIRE aircraft \citep{Forssell2005} on the form 
\begin{equation} 
\dot{x} = Ax + Bu, \quad y = Cx
\label{eq:sys} 
\end{equation}
where $x = \begin{bmatrix} \alpha & q \end{bmatrix}^T$ and $\alpha$ is the angle of attack and $q$ is the pitch rate of the aircraft, see Figure~\ref{fig:aircraft}, and $u$ is the elevator control surface deflection.
The matrices $A$ and $B$ vary when the c.g. shifts from its most forward position to its most aft (backward) position. With the c.g. in the most forward position the matrices are
\begin{equation} 
A = \begin{bmatrix} -1.453 & 0.9672 \\ 5.181 & -1.639 \end{bmatrix}, \quad B = \begin{bmatrix} 0.4467 \\ 34.79 \end{bmatrix}
\label{eq:sys_fwd}
\end{equation}
and when the c.g. is in the most aft position
\begin{equation} 
A = \begin{bmatrix} -1.45 & 0.9673 \\ 15.08 & -1.414 \end{bmatrix}, \quad B = \begin{bmatrix} 0.4461 \\ 31.77 \end{bmatrix}
\label{eq:sys_bwd}
\end{equation}
From this we can see that the force equation (first row of the matrices) is almost unaffected by the c.g. shift while the moment equation is largely affected by the shift in c.g. To stress the adaptive controller as much as possible we have designed the baseline controller for the most forward c.g. case and then simulate the total system with the model of the most aft c.g. case.

The baseline controller consist of an LQ feedback term, a static feed forward term to get a static gain of one between the reference and the output (angle of attack) and finally a integral part which integrates the error between the output and the nominal closed loop response (without the integral part). The baseline control signal is thus
\[ u_{bl} = -Kx + Fr + \int (y - y_{ref}) dt \]
where $y = \alpha$ and $y_{ref} = C(sI-A+BK)^{-1}BFr$. The baseline controller is designed using the matrices $A$ and $B$ from \eqref{eq:sys_fwd} but the $B$ matrix is simplified by setting the element in the force equation to zero, i.e., assuming that the control surface deflection do not generate any lift force but only moment. This approximation is not necessary at this stage but will have some nice implications in the adaptive design.

\begin{figure}[hbt]
  \centering
  \includegraphics[width=0.6\textwidth]{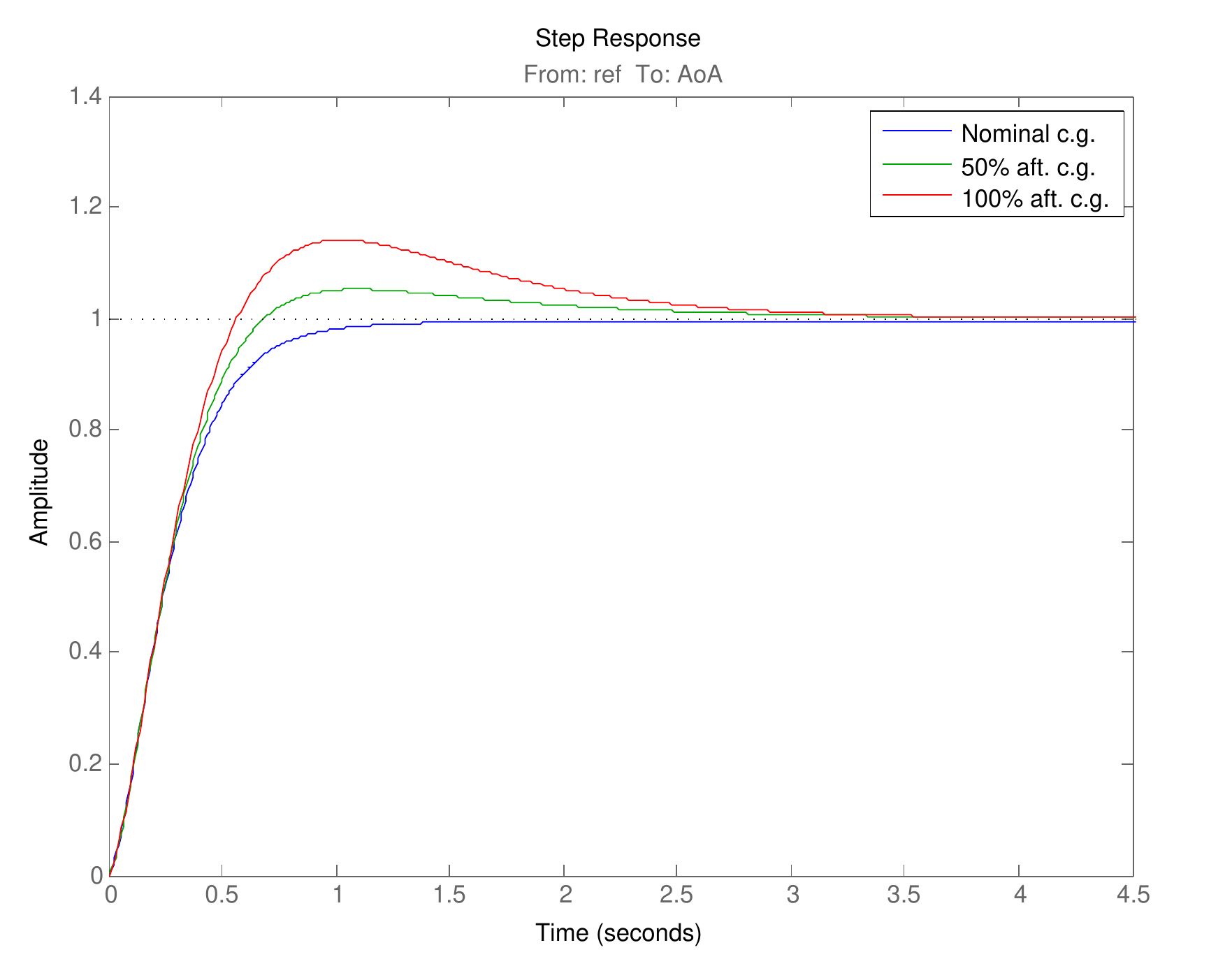}
  \caption{Step response of the closed loop system with nominal controller for different c.g positions}
  \label{fig:nom_response}
\end{figure}

In Figure~\ref{fig:nom_response} the response of the closed loop system with the nominal controller and different c.g. positions is shown. We can see that the response is good when the c.g. is at its nominal position (the blue line) but as the c.g. is moving backwards (green and red lines) there is a large overshoot. It is this overshoot that we want to minimize with an adaptive augmentation without destroying the nominal performance of the baseline controller.

\section{Robust MRAC design}
In Model Reference Adaptive Control (MRAC) one compare the output ($y$) of the closed loop system with that from a reference model
($y_m$). Then the controller parameters are updated such that the closed loop system response is as close as possible to that of the
reference system. The parameter update can be done in several different ways, e.g., by using the \emph{MIT-rule} or by using
Lyapunov stability theory. In this project we have chosen to use Lyapunov stability theory to derive the update laws for the controller
parameters. This is mainly due to the theoretical stability guarantees that comes with the method. In this report we will only briefly
describe the Lyapunov design process. For more information on the theoretical background of MRAC and the MIT and Lypunov update rules we refer the reader to the books of \cite{Astrom2008,Ioannou2012,Lavretsky2013}. 

In the MRAC design technique that we have adopted the uncertain system is modeled as
\begin{equation}
  \label{eq:unc_sys}
  \dot{x} = Ax + B\Lambda (u + \theta^T\phi(x) ) 
\end{equation}
where $A$ is an unknown matrix, $\Lambda$ is an unknown diagonal matrix and $B$ is known. The vector $\theta$ is the unknown coefficients of the general nonlinear function $\theta^T\phi(x)$ where $\phi(x)$ is a set basis functions. The aim of the adaptive controller is to have the system \eqref{eq:unc_sys} follow a reference model
\begin{equation}
  \label{eq:ref_sys}
  \dot{x} = A_m x + B_m r
\end{equation}
as close as possible with the use of the control signal $u = u_{ad} = -\hat{K}_x x + \hat{K}_r r$. This is only possible if there exist a set of ideal controller parameters $K_x^*$ and $K_r^*$ such that
\[ A - B\Lambda K_x^* = A_m, \quad B\Lambda K_r^* = B_m \]
These are the so called \emph{model matching conditions}.

Unfortunately the uncertainties in the model \eqref{eq:sys} due to the c.g. variations does not fulfil the model matching conditions, i.e., we can not use \eqref{eq:unc_sys} applied to system \eqref{eq:sys} to model the uncertainties. Instead we consider a model in the new state variable $z = [ \alpha \;\; \dot{\alpha}]^T$ which, if we use the same approximation of the $B$ matrix as used in the nominal controller, is a simple linear transformation of the original states
\[ z = Tx = \begin{bmatrix} 1 & 0 \\ a_{11} & a_{12} \end{bmatrix}x \]
and we get the following model of the uncertain system
\begin{equation}
  \label{eq:unc_sys_mod}
  \dot{z} = \underbrace{\begin{bmatrix} 0 & 1 \\ \tilde{a}_{21} & \tilde{a}_{22} \end{bmatrix}}_{\tilde{A}} z + \underbrace{\begin{bmatrix} 0 \\ \tilde{b}_2 \end{bmatrix}}_{\tilde{B}} u
\end{equation}
For this system the model matching conditions are fulfilled and we can model it with the structure \eqref{eq:unc_sys}, choosing $\Lambda = \lambda$ (scalar) and $\theta^T\phi(x) = 0$.  

The error between the closed loop system and the reference model can be written, using the model matching conditions and crudely ignoring the integral term in the nominal controller, as
\begin{equation}
  \label{eq:error}
  \dot{e}(t) =\dot{z}(t) - \dot{z}_m(t)   = \tilde{A}_me(t) - \tilde{B}^0\lambda\Delta\hat{K}_zz + \tilde{B}^0\lambda\Delta\hat{K}_r r
\end{equation}
where $\tilde{A}_m = TA_mT^{-1} = T(A-BK)T^{-1}$ and $\tilde{B}_m = TB_m = TBF$.

Using the Lyapunov function candidate
\begin{equation}
  \label{eq:lyapunov_cand}
  V(e,\Delta K_x,\Delta K_r) = \half \left(e^TPe + \frac{\abs{\lambda}}{\gamma_z} \Delta K_z \Delta K_z^T + \frac{\abs{\lambda}}{\gamma_r} \Delta K_r^2 \right)
\end{equation}
and differentiate w.r.t. time, using \eqref{eq:error}, we obtain
\begin{align*}
  \dot{V} & = \half \left( \dot{e}^TPe + e^TP\dot{e} 
 + \frac{\abs{\lambda}}{\gamma_z} \Delta \dot{K}_z \Delta K_z^T + \frac{\abs{\lambda}}{\gamma_z} \Delta K_z \Delta \dot{K}_z^T 
 + 2\frac{\abs{\lambda}}{\gamma_r} \Delta K_r \Delta \dot{K}_r \right)\\
& = \half e^T\left(\tilde{A}_m^TP + P\tilde{A}_m\right)e + \frac{\abs{\lambda}}{\gamma_z}\Delta \dot{K}_z \Delta K_z^T - \lambda e^TP\tilde{B}^0\Delta K_z z
+ \frac{\abs{\lambda}}{\gamma_r} \Delta \dot{K}_r \Delta K_r + \lambda e^TP\tilde{B}^0\Delta K_r r\\
& =  - \half e^TQe + \left(\frac{\abs{\lambda}}{\gamma_z}\Delta \dot{K}_z - \lambda e^TP\tilde{B}^0\right) \Delta K_z^T + \left( \frac{\abs{\lambda}}{\gamma_r} \Delta \dot{K}_r + \lambda e^TP\tilde{B}^0 \right) \Delta K_r 
\end{align*}

If the derivative is negative then \eqref{eq:lyapunov_cand} is a valid Lyapunov function and the closed loop system is stable. A negative derivative is obtained if $\tilde{A}_m^TP + P\tilde{A}_m = -Q$ for some $Q > 0$ and if we select the adaptive controller gains as
\begin{subequations} \label{eq:mrac_update}
\begin{align}
  \dot{K}_z &= \gamma_z\text{sgn}(\lambda) e^TP\tilde{B}^0z^T \\
  \dot{K}_r & = - \gamma_r\text{sgn}(\lambda) e^TP\tilde{B}^0r
\end{align}
\end{subequations}

To be able to cope with disturbances, sensor noise and to not interfere with the nominal controller we need to add some additional ingredients to the update laws \eqref{eq:mrac_update}. To not have the parameters drift due to noise we add a limit on the parameters in the form of a projection operator \citep{Ioannou2012}. In addition to this we also add a dead zone for small errors, $e(t)$. This also reduces the sensitivity to noise but additionally it makes the adaptive controller not to interfere with the performance of the nominal controller \citep{Lavretsky2013}. With these modifications the adaptive laws become
\begin{subequations} \label{eq:robust_mrac_update}
\begin{align}
  \dot{K}_z &= \begin{cases} \text{Proj} \left(K_z, \gamma_z\text{sgn}(\lambda) e^TP\tilde{B}^0z^T \right) & \norm{e(t)}{} > \epsilon \\ 0 & \norm{e(t)}{} \leq \epsilon \end{cases} \\
  \dot{K}_r & = \begin{cases} \text{Proj} \left(K_r, - \gamma_r\text{sgn}(\lambda) e^TP\tilde{B}^0r \right) & \norm{e(t)}{} > \epsilon \\ 0 & \norm{e(t)}{} \leq \epsilon \end{cases}
\end{align}
\end{subequations}

Finally, the adaptive augmentation control used is $u_{ad} = - K_z z +K_r r$, where $K_z$ and $K_r$ are updated according to  \eqref{eq:robust_mrac_update}.

\begin{figure}[hbt]
  \centering
  \includegraphics[width = 0.6\textwidth]{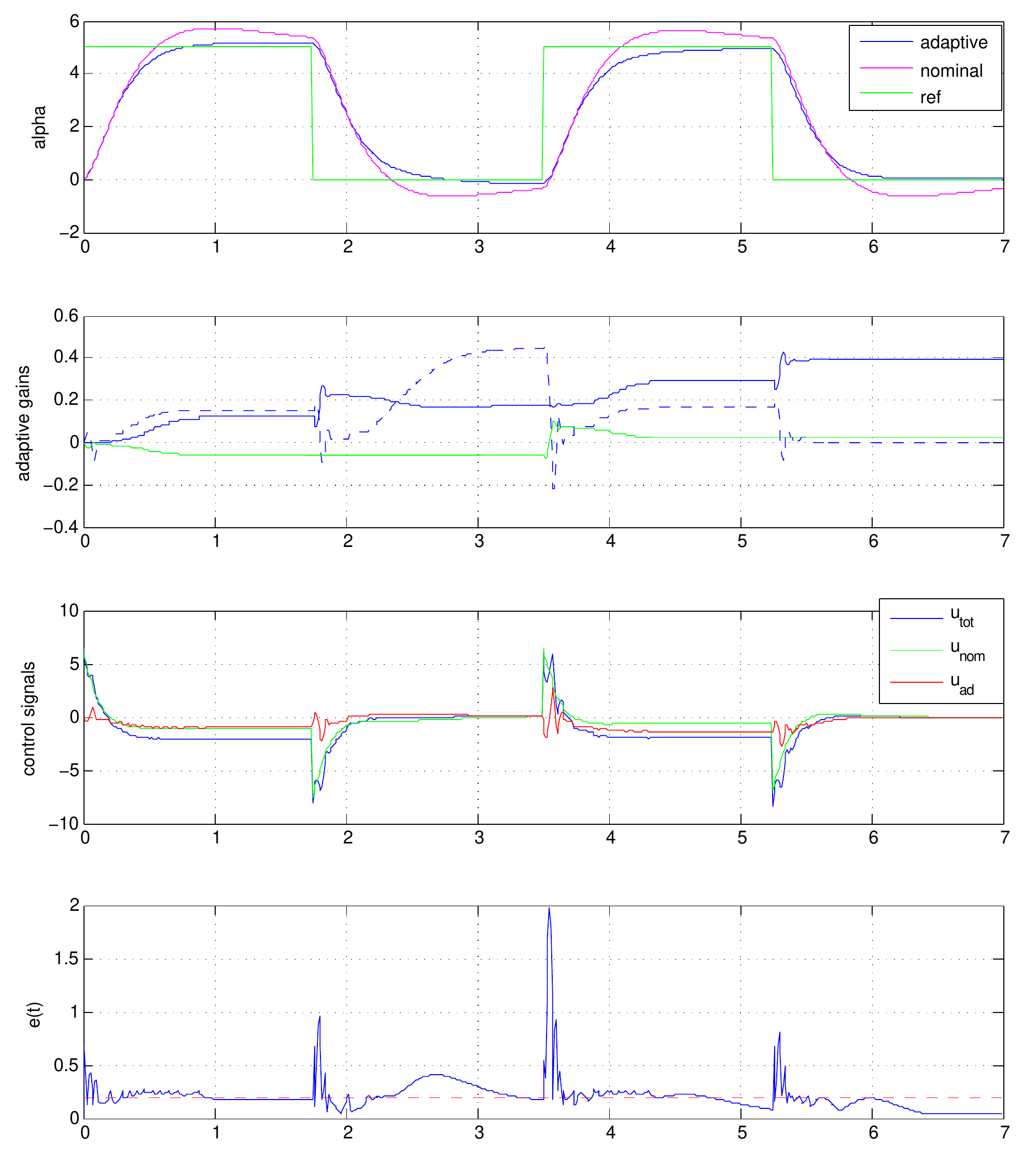}
  \caption{Simulation of the closed loop system with both baseline controller and adaptive augmentation. The simulated system is the aircraft with the c.g. in its most aft position.}
  \label{fig:simulation}
\end{figure}

We have validated the design by simulating the system \eqref{eq:sys} with the c.g. in its most aft position, i.e., with the use of matrices \eqref{eq:sys_bwd}. This system has been controlled with the baseline controller designed for the case with matrices \eqref{eq:sys_fwd} and with the adaptive augmentation. In Figure~\ref{fig:simulation} we see the result of the simulation.

In the first part of the figure we see the angle of attack response of the closed loop adaptive system (blue) compared to using only the baseline controller (magenta). We can see that already in the beginning, when the adaptive controller has not converged, the performance is better with the adaptive augmentation. At the end of the simulation, when the adaptive controller has ``converged'' we can see that the response is really good.

In the second part of the figure we can see the evolution of the adaptive parameters. The blue solid line is the angle of attack feedback term, the blue dashed line is the pitch rate feedback and the green line is the reference feedforward term. An interesting observation is that at the end when the performance is best the pitch rate feedback and the feedforward gains are almost zero. Therefore it might be interesting to try and use only an angle of attack feedback term in the adaptive augmentation. 

In the third part of the figure the baseline control signal, adaptive augmentation control signal as well as the total control signal is shown. Here we can see that the adaptation causes ripples on the control signal. The magnitude and frequency of the ripples is dependent of the tuning but it is difficult to remove it completely.

In the last part of the figure we plot the 2-norm of the model following error, $e(t) = z(t) - z_m(t)$ together with the dead zone level, $\epsilon$ (red dashed line). 


\section{Open issues and future work}
Even though the validating simulation looks good there are still some issues that needs to be further investigated before we can conclude that adaptive augmentation of this form is a good methodology to use for c.g. adaptation in aircrafts.

The first issue is that even though the Lyapunov design method shall have guaranteed stability it was possible to make the closed loop instable for different tuning and reference signals.

Secondly the adaptive control laws are not scale invariant to a large range of reference signals. A tuning that works good for small reference signal values makes the closed loop unstable for large reference signals and if the adaptive laws are tuned for large reference signals then there will be poor performance for small signals. Additionally the dead zone level, $\epsilon$, is dependent on the signal levels. This phenomenon is pointed out in \cite{Astrom2008} and both they and \cite{Ioannou2012} suggest a normalisation scheme for the update laws. However the normalised adaptive laws in both \cite{Astrom2008} and \cite{Ioannou2012} are designed in a transfer function framework and we have found no equivalent in the state space setting. \cite{Lavretsky2013} suggest as an alternative to use an integral feedback term instead of the reference feedforward term in the adaptive control signal.

The most important issues to continue working on is probably the scale invariance of the controller since this affects both stability and the tuning of the dead zone. The idea is to try and find a suitable way to normalize the adaptive laws in the state space setting. The possibility to use only an angle of attack feedback as adaptive augmentation is also a very interesting idea to investigate further.

\bibliographystyle{abbrvnat}
\bibliography{library}

\begin{thebibliography}{4}
\providecommand{\natexlab}[1]{#1}
\providecommand{\url}[1]{\texttt{#1}}
\expandafter\ifx\csname urlstyle\endcsname\relax
  \providecommand{\doi}[1]{doi: #1}\else
  \providecommand{\doi}{doi: \begingroup \urlstyle{rm}\Url}\fi

\bibitem[{\AA}str{\"{o}}m and Wittenmark(2008)]{Astrom2008}
K.~J. {\AA}str{\"{o}}m and B.~Wittenmark.
\newblock \emph{{Adaptive Control}}.
\newblock Dover, 2 edition, 2008.
\newblock ISBN 978-0-486-46278-3.

\bibitem[Forssell and Nilsson(2005)]{Forssell2005}
L.~Forssell and U.~Nilsson.
\newblock {ADMIRE The Aero-Data Model In a Research Environment Version 4.0,
  Model Description}.
\newblock Technical Report December, Swedish Defence Research Agency, 2005.

\bibitem[Ioannou and Sun(2012)]{Ioannou2012}
P.~A. Ioannou and J.~Sun.
\newblock \emph{{Robust Adaptive Control}}.
\newblock Dover, 2012.
\newblock ISBN 978-0-486-49817-1.

\bibitem[Lavretsky and Wise(2013)]{Lavretsky2013}
E.~Lavretsky and K.~A. Wise.
\newblock \emph{{Robust and Adaptive Control With Aerospace Applications}}.
\newblock Springer, 1 edition, 2013.
\newblock ISBN 978-1-4471-4395-6.

\end{thebibliography}

\end{document}